\documentclass[3p]{elsarticle}
\usepackage{graphics}
\usepackage{amssymb}
\usepackage{bm,color,ulem}
\usepackage{hyperref}
\usepackage{comment}

\journal{Physica A}
\begin{document}
\begin{frontmatter}

\title{Entropic measures of joint uncertainty: effects of lack of majorization}

\author[ucm]{Alfredo Luis \corref{ca}}
\ead{alluis@ucm.es}
\cortext[ca]{Corresponding author}

\author[ulp]{Gustavo Mart\'{\i}n Bosyk}
\author[ulp,lgi]{Mariela Portesi}

\address[ucm]{Departamento de \'{O}ptica, Facultad de Ciencias F\'{\i}sicas, Universidad
Complutense, 28040 Madrid, Spain}
\address[ulp]{Instituto de F\'{\i}sica La Plata (IFLP), CONICET, and Departamento de
F\'{\i}sica, Facultad de Ciencias Exactas, Universidad Nacional de La Plata, Casilla de
Correo 67, \\ 1900 La Plata, Argentina}
\address[lgi]{Laboratoire Grenoblois d'Image, Parole, Signal et Automatique (GIPSA-Lab, CNRS),
11 rue des Math\'ematiques, 38402 Saint Martin d'H\`eres, France}

\begin{abstract}
We compute R\'{e}nyi entropies for the statistics of a noisy simultaneous observation of two
complementary observables in two-dimensional quantum systems. The relative amount of uncertainty between two states depends on the uncertainty measure used. These results are not reproduced by a more standard duality relation.
We show that  these behaviors are consistent with the lack of majorization relation between the corresponding statistics.
\end{abstract}


\end{frontmatter}

\section{Introduction}

Historically, the joint uncertainty of pairs of observables has been mostly addressed
in terms of the product of their variances. Nevertheless, there are situations where such
formulation is not satisfactory enough~\cite{pvar}, thus alternative approaches have
been proposed, mainly in terms of diverse entropic measures~\cite{EURs1,EURs2,ZPV,PhysA} (see also the reviews in~\cite{Reviews}).
In this work we consider in particular the so-called R\'{e}nyi entropies~\cite{Ry} and
the corresponding entropic uncertainty relations, for the statistics associated to two complementary observables~\cite{BPP}.
There has been an increasing activity to obtain different and improved entropic uncertainty relations not only for foundational reasons but also for the different applications in quantum information problems (a non-exhaustive list includes information-theoretic formulation of error--disturbance relations~\cite{error}, connection with duality relations~\cite{Coles2014} and nonlocality~\cite{Tomamichel2013}, entanglement detection~\cite{ent}, EPR-steering inequalities~\cite{EPRsteering}, quantum memory~\cite{Berta2010}, and security of quantum cryptography protocols~\cite{crypto}).
Also, entropic uncertainty relations have a deep connection with the majorization of statistical
distributions~\cite{HP,FGG}, which has been already applied to examine uncertainty of
thermal states~\cite{CRP} (this is closely related to the idea of mixing character~\cite{Ruch}).

However, previous works~\cite{ZPV,BPP} have shown that entropic uncertainty relations may lead
to unexpected results, derived from the fact that the amount of uncertainty for a pair of
observables depends on the uncertainty measure used. This is quite natural; actually,
one of the benefits of using entropic measures is that they adapt to asses different operational
tasks. Nevertheless, one may find it surprising  that different measures  lead to opposite conclusions
in entropic relations: this is, that the states of maximum uncertainty for one measure are the
minimum uncertainty states for the other measure, and vice versa.

In this regard, the aim of this work is twofold. On the one hand, we show that
these unexpected behaviors
are fully compatible with the
lack of majorization relation between the corresponding statistics. This connection
holds because entropic measures are monotone with respect to majorization.
Thus, such surprising entropic results are not tricky features of entropic measures, but
may have a deeper meaning that is actually overlooked by more popular measures
of uncertainty or complementarity. On the other hand, we extend the application of
entropic measures to the statistics of a simultaneous joint observation of two complementary
observables in the same system realization~\cite{AG,WMM,BAB}. This setting of complementarity
in practice provides a rich arena to examine the interplay between entropic measures
and majorization. The simultaneous measurement provides a true joint classical-like
probability distribution that enables alternative assessments of joint uncertainty, different
from the ones given by the product of individual statistics, either intrinsic or of
operational origin.

For simplicity we address these issues in the simplest quantum system described
by a state in a two-dimensional Hilbert space. This comprises very relevant practical
situations such as the path--interference complementarity in two-beam interference experiments.
This allows us to contrast the performance of entropic measures with respect to more standard
descriptions of complementarity ~\cite{duality1,duality2,BPHP}.

The paper is organized as follows: in Sec.~\ref{simult:sec} we introduce the discussion on
statistics of simultaneous measurements for spin $\frac{1}{2}$ observables. Sec.~\ref{entropicunc:sec}
exhibits noticeable results for entropic quantities, and an explanation for that behavior is
given in Sec.~\ref{majorization:sec}. In Sec.~\ref{duality:sec}, a duality relation for complementarity 
is analyzed and compared with the entropic results. Finally, some concluding remarks are outlined 
in Sec.~\ref{conclu:sec}.

\section{Statistics and simultaneous measurements}
\label{simult:sec}

Let us consider two complementary observables represented by the Pauli spin matrices
$\sigma_x$ and $\sigma_z$. In practical terms they may represent phase and path, 
respectively, in two-beam interference experiments. The system state is described by a
density matrix operator acting on the Hilbert space ${\cal H}_S$ that in Bloch representation
acquires the form \ $\rho = \frac{1}{2} \left ( I + \bm{s} \cdot \bm{\sigma} \right )$, where $I$
is the identity matrix, $ \bm{\sigma}$ represents the three Pauli matrices, and $\bm{s} =
\mathrm{Tr} (\rho \, \bm{\sigma})$ is a three-dimensional Bloch vector with $|\bm{s} | \leq 1$.
The modulus $| \bm{s} |$ expresses the degree of purity of the state as $\mathrm{Tr} ( \rho^2 )
= \frac 12 (1+ | \bm{s} |^2 )$,  \ being $|\bm{s} |=1$ in the case of a pure state. We make use
of the Bloch-sphere parametrization:
\begin{equation}
\label{pss}
s_x = |\bm{s} | \sin \theta \cos \varphi,  \quad s_y = |\bm{s} | \sin \theta \sin \varphi, \quad
s_z = |\bm{s} | \cos \theta .
\end{equation}
The \textit{intrinsic statistics} for the observables $\sigma_x$ and $\sigma_z$ are
\begin{equation}
\label{mis}
p^X_j = \frac{1}{2} \left ( 1 + j \, s_x \right )\textit
\quad \mbox{and} \quad
p^Z_k = \frac{1}{2} \left ( 1 + k \, s_z \right ) ,
\end{equation}
with $j = \pm 1$ and $k= \pm 1$.

The \textit{simultaneous measurement} of noncommuting observables requires involving auxiliary
degrees of freedom, usually referred to as apparatus. In our case we consider an apparatus
described by a two-dimensional Hilbert space $ {\cal H}_A$. The measurement performed in
${\cal H}_A$ addresses that of $\sigma_z$, 
while $\sigma_x$ is measured directly on the system space ${\cal H}_S$. The system--apparatus
coupling transferring information about $\sigma_z$ from the system to the apparatus is arranged
via the following unitary transformation acting on $ {\cal H}_S \otimes {\cal H}_A$,
\begin{equation}
\label{U}
U = | + \rangle \langle + | \otimes \ U_+ + | - \rangle \langle - | \otimes \ U_-,
\end{equation}
where $U_\pm$ are unitary operators acting solely on ${\cal H}_A$, while $|\pm \rangle$ are the
eigenstates of $\sigma_z$ with corresponding eigenvalues $\pm 1$.
For simplicity the initial state of the apparatus, $| a \rangle \in {\cal H}_A$, is
assumed to be pure, so that the system--apparatus coupling leads to
\begin{equation}
\label{a_pm}
U | + \rangle | a \rangle \rightarrow  \ | + \rangle \ | a_+ \rangle , 	\quad
U | - \rangle | a \rangle \rightarrow \ | - \rangle \ | a_- \rangle,
\end{equation}
where the states $| a_\pm \rangle = U_\pm | a \rangle \in {\cal H}_A$ are not orthogonal
in general, with $\cos \delta = \langle a_+ | a_- \rangle$ assumed to be a positive real
number with  $0 \leq \delta \leq \pi/2$, without loss of generality. The measurement in ${\cal H}_A$ introducing minimum
additional noise is given by projection on the orthogonal vectors $| b_\pm \rangle$ \
(see Fig.~1):
\begin{eqnarray}
\label{m12}
 | b_+ \rangle &=& \frac{1}{\cos \phi} \left ( \cos \frac{\phi}{2}
\, | a_+ \rangle - \sin \frac{\phi}{2} \, | a_- \rangle \right ) ,  \nonumber \\
 | b_- \rangle &=& \frac{1}{\cos \phi} \left ( - \sin \frac{\phi}{2} \, | a_+ \rangle
+ \cos \frac{\phi}{2} \, | a_- \rangle \right ) ,
\end{eqnarray}
where $\phi = \frac{\pi}{2} - \delta$. The added noise is minimum in the sense that the Euclidean distance
between $p^Z$ and the marginal probability $\tilde{p}^Z$ (defined in Eq.~(\ref{ms}) below) is minimum \cite{Luis2013}. It is worth noting that
there is a deep connection of this measurement  with the problem of state discrimination between
two nonorthogonal states, such as $| a_\pm \rangle$ (see e.g.~\cite{sd}).

\begin{figure}
\begin{center}
\includegraphics[scale=0.2]{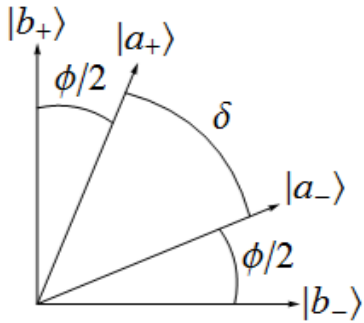}
\end{center}
\caption{Schematic representation of the states $|a_\pm \rangle$ and $| b_\pm \rangle$, given in
Eqs.~(\ref{a_pm}) and~(\ref{m12}) respectively.}
\end{figure}

The \textit{joint statistics} for the simultaneous measurement of $\sigma_x$ acting on
${\cal H}_S$ and of $\sigma_z$ addressed by the orthogonal vectors $| b_\pm \rangle$ in ${\cal H}_A$ is
\begin{equation}
\label{js}
\tilde{p}^{X,Z}_{j,k} = \frac{1}{4} ( 1 + j\, s_x \cos\delta + k\, s_z \sin\delta ) ,
\end{equation}
where $j=\pm 1$ represents the outcomes of the $\sigma_x$ measurement,
and $k=\pm 1$ those of the $\sigma_z$ measurement.
The \textit{marginal statistics} for both observables are
\begin{equation}
\label{ms}
\tilde{p}^X_j = \frac{1}{2} ( 1 + j\, s_x \cos\delta ) \quad \mbox{and}
\quad
\tilde{p}^Z_k = \frac{1}{2} ( 1 + k\, s_z \sin\delta ) .
\end{equation}
When contrasted with the intrinsic statistics~(\ref{mis}) we get that the
observation of $\sigma_x$ is exact for $\delta= 0$, while the observation of
$\sigma_z$ is exact for $\delta= \frac{\pi}{2}$. For $\delta = \frac{\pi}{4}$, the extra
uncertainty introduced by the unsharp character of the simultaneous observation
is balanced between observables.

The expressions given above are valid for any system state $\rho$. However for the sake
of simplicity, and given that we focus on the observables $\sigma_{x}$ and $\sigma_{z}$,
we  will frequently particularize to the set $S$ of states with Bloch vector $\bm{s}$ lying in the $XZ$
plane, this is, for $s_y =0$ and $\varphi =0$.

\section{Entropic uncertainty assessments}
\label{entropicunc:sec}

\subsection{R\'enyi entropies and entropic uncertainty relations}

We make use of generalized entropies  to quantify the uncertainty
(or ignorance) related to a probability distribution. Let $p=(p_1, \ldots, p_N)$ be
the statistics of some observable with $N$ outcomes, then the R\'enyi entropy~\cite{Ry} of order
$\alpha$ reads
\begin{equation}
\label{Renyient}
R_{\alpha} (p) = \frac{1}{1-\alpha} \ln \left ( \sum_{i=1}^N p_i^{\,\alpha} \right ),
\end{equation}
where $\alpha \geq 0$ is the so-called entropic index\footnote{The entropic index $\alpha$ plays the role of a magnifying glass: for $\alpha < 1$ the contribution of the terms in the sum in~(\ref{Renyient}) becomes more uniform than in the case $\alpha = 1$; whereas for $\alpha > 1$, the leading probabilities of the distribution are stressed.
}.
Notice that Shannon entropy~\cite{Shannon1948}, $- \sum_i p_i \ln p_i$, is recovered in the limiting case
$\alpha \rightarrow 1$.
For vanishingly small $\alpha$, one has $R_0 = \ln||p||_0$, where $||p||_0$ is the number of nonzero components
of the statistics, whereas for arbitrary large $\alpha$, $R_\infty = -\ln \max_i p_i $ only takes into account the greatest component
of the statistics and is known as min-entropy, due to the nonincreasing property of $R_\alpha$
versus $\alpha$ for a given $p$.

An important property of R\'enyi entropies is related to majorization (see e.g.~\cite{MarshallBook}).
It is said that a statistics $p$ majorizes a statistics $p^\prime$, denoted as $p^\prime \prec p$,
if after forming with $p$ an $N$-dimensional vector with components in
decreasing order ($p_1 \geq p_2 \geq \ldots \geq p_N$)
and similarly with $p^\prime$,
the inequalities \  $\sum_{i=1}^k p^\prime_i \leq \sum_{i=1}^k p_i$ \ are fulfilled for all $k=1,2,\ldots,N-1$ and
$\sum_{i=1}^N p^\prime_i = \sum_{i=1}^N p_i=1$.
Notice that, by definition, the values of $R_\alpha$ remains unaltered under rearrangement of the probability vector.
The R\'enyi entropies are order-preserving or Schur-concave functions, this means:
\begin{center} If \ $p^\prime \prec p$, then \ $R_{\alpha} (p^\prime ) \geq R_{\alpha} (p)$ for any $\alpha \geq 0$.
\end{center}
However, majorization is a relation of partial order, so that there are distributions that cannot be compared.
We will see that the behaviours that we reporte in the next section are consistent with
lack of majorization.

The Schur-concavity property allows to show that R\'enyi entropies are lower and upper bounded: \ since
$(\frac{1}{N} , \ldots , \frac{1}{N}) \prec p \prec (1, 0, \ldots, 0)$ \ then, for every $\alpha$, one has \ $0 \leq R_{\alpha} (p) \leq \ln N$,
where the bounds are attained if and only if the corresponding majorization relations reduce to equalities (i.e., for the completely certain situation and the fully random, respectively).

Other relevant property of $R_{\alpha}$ is additivity, that is, for the product of two statistics $p$ and $q$ one has
\begin{equation}
 R_{\alpha} (p \, q) = R_{\alpha} ( p ) + R_{\alpha} ( q ).
 \end{equation}
However R\'enyi entropies do not satisfy, in general, subadditivity and concavity properties~\footnote{Subadditivity is valid only for $\alpha=0$ and $\alpha=1$~\cite[p.149]{Aczelbook}; concavity holds for all $\alpha \in [0,1]$, whereas for $\alpha >1 $ the concavity is held up to an index $\alpha^*$ that depends on $N$~\cite[p.57]{BengtssonBook}.}.

Following Refs.~\cite{BPP}, it can be seen that the R\'enyi entropic uncertainty relations corresponding to
 the intrinsic statistics~(\ref{mis}) are:
\begin{equation}
 R_{\alpha} (p^X p^Z) \geq   \left\{
                                \begin{array}{ll}
                                  \ln 2 & \hbox{if } 0 \leq \alpha \leq \alpha_I \\
                                  \frac{2}{1-\alpha} \, \ln\left[\left(\frac{1+\frac{1}{\sqrt2}}{2} \right)^\alpha +
                                  \left(\frac{1-\frac{1}{\sqrt2}}{2} \right)^\alpha \right] & \hbox{if } \alpha > \alpha_I,
                                \end{array}
                              \right.
 \end{equation}
where $\alpha_I \approx 1.43$. There are two subsets of states within the set $S$ that compete to be the \textit{minimum uncertainty states}
(as well as those of maximum uncertainty), depending on the value of the entropic index used. We refer
to them as \textit{extreme} and \textit{intermediate} states:
\begin{itemize}
\item
Extreme states are eigenstates of $\sigma_x$ \ or \ $\sigma_z$. These are pure
states with $\theta= m \frac{\pi}{2}$ for integer $m$, then $s_x = \pm 1, s_y =0= s_z$, \ or \ $s_x =0=s_y, s_z = \pm 1$. They present full certainty for one observable, and complete uncertainty for the other one.
\item
Intermediate states are eigenstates of \ $\sigma_x \pm \sigma_z$. These are
 pure states with $\theta= (2 m +1) \frac{\pi}{4}$ for
integer $m$, then $s_x = \pm s_z$, $s_y=0$ and $s_z = \pm \frac{1}{\sqrt{2}}$.
They have essentially the same statistics for both complementary observables so they can
be considered as a finite-dimensional counterpart of the Glauber coherent states.
\end{itemize}

More generally, one has the following mixed versions of extreme and  intermediate states, respectively,
\begin{equation}
\label{mix}
\rho^X _{\mathrm{ex}} = \frac{1}{2} \left ( I \pm | \bm{s} | \sigma_{x} \right ), \quad
\rho^Z_{\mathrm{ex}} = \frac{1}{2} \left ( I \pm | \bm{s} | \sigma_{z} \right ), \quad
\rho_{\mathrm{in}} = \frac{1}{2} \left [ I \pm \frac{| \bm{s} |}{\sqrt{2}} \left (  \sigma_x  \pm \sigma_z
\right ) \right ],
\end{equation}
with $| \bm{s} |$ expressing the degree of purity.

\subsection{Extreme versus intermediate states}
\label{intvsextr:sec}

In order to assess the uncertainty related to $\sigma_x$ and $\sigma_z$, we compute the R\'enyi
entropies of the joint statistics $\tilde{p}^{X,Z}$~(\ref{js}) and of the product of marginal statistics
$\tilde{p}^X\tilde{p}^Z$~(\ref{ms}), for any given value of the entropic index $\alpha$, as functions
of $\theta$ within the set $S$ of states.
These quantities are calculated for balanced
measurement, $\delta = \frac{\pi}{4}$. We also take into account the R\'enyi entropies of the product of
intrinsic statistics $p^X p^Z$~(\ref{mis}). For the sake of clarity and to simplify comparisons, we
mostly focus on normalized quantities of the form
\begin{equation}
\label{norm}
R^{\mathrm{norm}}_{\alpha}(p) = \frac{R_{\alpha}(p) - R_{\alpha,\mathrm{min}}(p)}{R_{\alpha,\mathrm{max}}(p)-
R_{\alpha,\mathrm{min}}(p)} ,
\end{equation}
where $R_{\alpha,\mathrm{max}}$ and $R_{\alpha,\mathrm{min}}$ are the maximum and minimum
values of $R_{\alpha}$, respectively, within the set $S$.

\begin{figure}
\begin{center}
\includegraphics[scale=0.30]{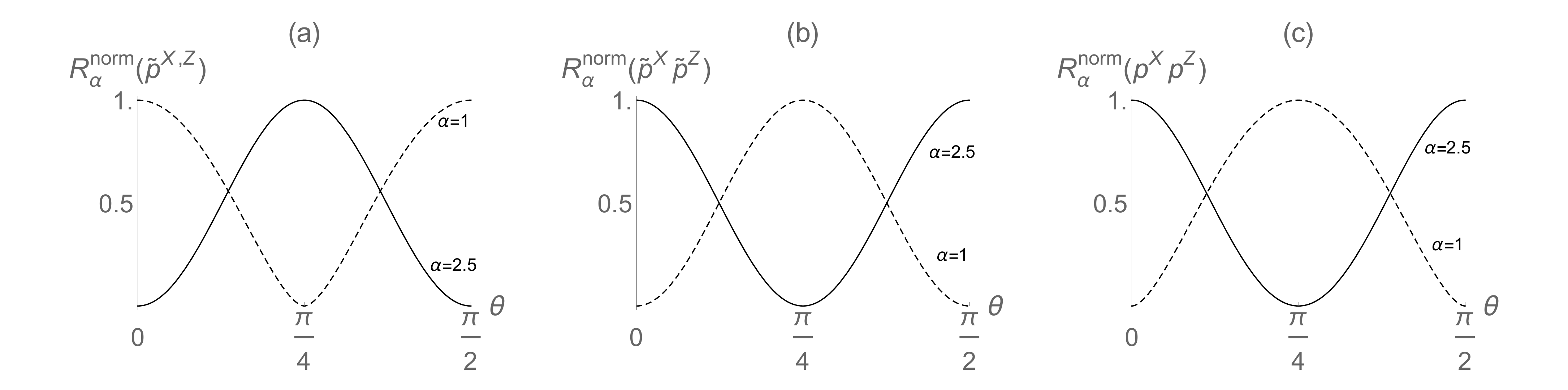}
\end{center}
\caption{Normalized R\'{e}nyi entropies~(\ref{norm}) of: \ (a)~the joint statistics $R^{\mathrm{norm}}_{\alpha} (\tilde{p}^{X,Z} )$,
(b)~the product of marginal statistics $R^{\mathrm{norm}}_{\alpha} (\tilde{p}^X \tilde{p}^Z )$, and (c)~the product
of intrinsic statistics $R^{\mathrm{norm}}_{\alpha} (p^X p^Z)$, for $\alpha=1$ (dashed lines) and $\alpha=2.5$ (solid
lines), as functions of $\theta$ for $\varphi=0$, $| \bm{s} | =1$ and $\delta = \frac{\pi}{4}$. The states corresponding to minimal uncertainty vary depending the entropic index used, within each statistical description.}
\label{Rentvstheta:fig}
\end{figure}

Figure~\ref{Rentvstheta:fig}(a) shows $R_{\alpha}^{\mathrm{norm}}(\tilde{p}^{X,Z})$ for $\alpha=1$
and $2.5$ as functions of $\theta$, taking $\varphi=0$ and $| \bm{s} | =1$ for simplicity.
We observe that for $\alpha=1$ the minimum uncertainty states are the intermediate states $\theta = \frac{\pi}{4}$,
whereas for $\alpha=2.5$ the minimum uncertainty states are the extreme states $\theta = 0$ or $\frac{\pi}{2}$.
The opposite happens for the product of marginal statistics $R_{\alpha} ( \tilde{p}^X \tilde{p}^Z )$, as illustrated in
Fig.~\ref{Rentvstheta:fig}(b) (that is, for $\alpha=1$ the minimum uncertainty states are the extreme states $\theta = 0$ or
$\frac{\pi}{2}$, whereas for $\alpha=2.5$ the minimum uncertainty states are the intermediate states $\theta = \frac{\pi}{4}$).
The latter result coincides with the conclusions derived from the intrinsic entropies $R_{\alpha} (p^X p^Z)$ as shown in
Fig.~\ref{Rentvstheta:fig}(c) (see also Refs.~\cite{BPP}).

\begin{figure}
\begin{center}
\includegraphics[scale=0.7]{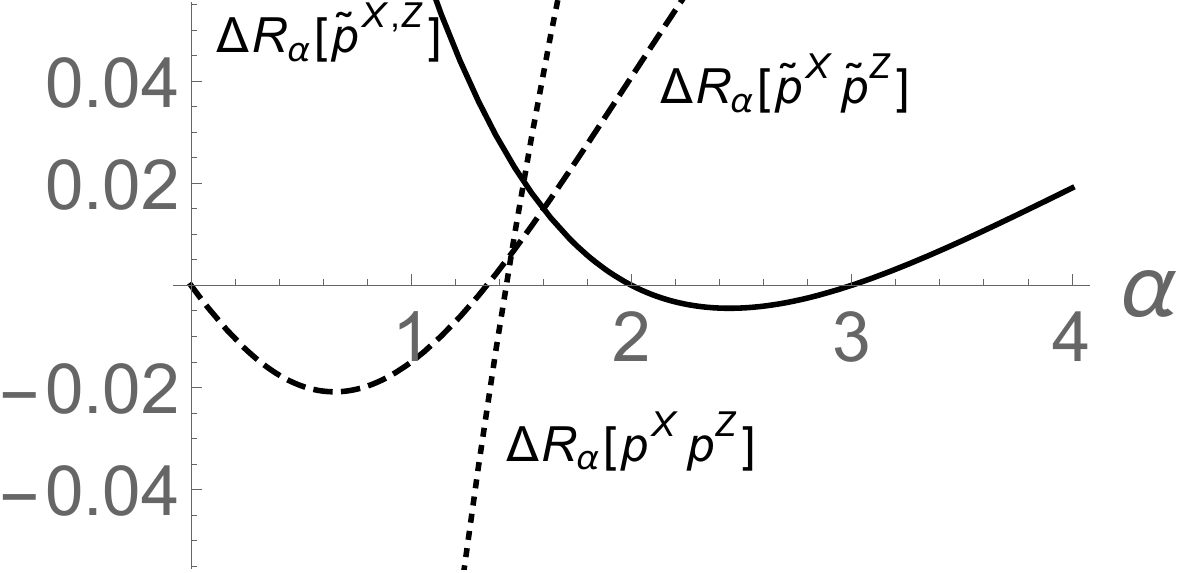}
\end{center}
\caption{Differences~(\ref{iec}) between R\'enyi entropies for intermediate and extreme states in the cases of joint statistics
$\Delta R_{\alpha} [\tilde{p}^{X,Z}]$ (solid line), product of marginals $\Delta R_{\alpha} [\tilde{p}^X \tilde{p}^Z]$
(dashed line), and the product of intrinsic statistics $\Delta R_{\alpha} [p^X p^Z] $ (dotted line), as functions of $\alpha$
for  $\varphi=0$, $| \bm{s} | =1$, and $\delta = \frac{\pi}{4}$. A negative value of
$\Delta R_{\alpha}$ means that extreme states give the minimum uncertainty, whereas a psotive value corresponds to
minimizing intermediate states. }
\label{extvsint:fig}
\end{figure}

In order to make more explicitly the extreme--intermediate competition for minimum uncertainty, we plot the
difference of R\'{e}nyi entropies between extreme and intermediate states as functions of the entropic index~$\alpha$ for $\varphi=0$ and $| \bm{s} | =1$, namely,
\begin{equation}
\label{iec}
\Delta R_{\alpha}[p] = R_{\alpha} (p_\mathrm{ext}) - R_{\alpha} (p_\mathrm{int}),
\end{equation}
where $p$ stand for the joint, product of marginals, and product of intrinsic statistics, considering balanced measurement $\delta = \frac{\pi}{4}$ (see Fig.~\ref{extvsint:fig}). $\Delta R_{\alpha}  <0$ implies that
extreme states are of minimum uncertainty while, on the contrary, $\Delta R_{\alpha} > 0$ implies that intermediate states
are the minimum uncertainty ones.

We observe that, for the joint statistics, $\Delta R_{\alpha} [\tilde{p}^{X,Z}] $ is negative if $\alpha \in (2,3)$, ­
thus there are two critical values of the entropic index at which the minimizer changes. On the other hand, for
the products of marginal and intrinsic statistics, $\Delta R_{\alpha} [\tilde{p}^X \tilde{p}^Z] $ and $\Delta
R_{\alpha} [p^X p^Z]$ change their sign at one critical value: $\alpha_M \approx 1.34$ in the former case
and $\alpha_I \approx 1.43$ in the latter; in both situations, the difference changes from negative to
positive as the entropy index increases.

Let us mention that similar results can be obtained by using the family of Tsallis entropies~\cite{Ts}, since there is a one-to-one correspondence between R\'enyi and Tsallis families of entropies.

\section{Majorization assessments}
\label{majorization:sec}

\subsection{Extreme and intermediate states are incomparable}
\label{ext&int maj:sec}

Let us call $\tilde{\lambda}=\tilde{p}^{X,Z}_{\textrm{ex}}$ and $\tilde{\mu}=\tilde{p}^{X,Z}_{\textrm{in}}$ 
the four-dimensional vectors obtained by arranging the values of $\tilde{p}^{X,Z}$ in decreasing
order, for extreme and intermediate states, respectively. After Eqs. (\ref{js}) and (\ref{mix}) we get for $\delta = \pi/4$:
$$\tilde{\lambda} = \frac{1}{4} \left(1+\frac{|\bm{s} |}{\sqrt{2}}, 1+ \frac{|\bm{s} |}{\sqrt{2}},  1- \frac{|\bm{s} |}
{\sqrt{2}},1- \frac{|\bm{s} |}{\sqrt{2}}\ \right) , $$
and
$$\tilde{\mu} = \frac{1}{4} \left(1+|\bm{s} |, 1, 1, 1- |\bm{s} | \right).$$
Thus for all $|\bm{s} | \neq 0$ we have clearly $\tilde{\mu}_1 > \tilde{\lambda}_1$ but  $\tilde{\mu}_1 + \tilde{\mu}_2  < \tilde{\lambda}_1 + \tilde{\lambda}_2 $
so that neither $\tilde{\lambda} \prec \tilde{\mu}$ nor $\tilde{\mu} \prec \tilde{\lambda}$. This shows that contradictions hold for pure as well as for
mixed states, while naturally the differences between the extreme and intermediate states are larger for larger $|\bm{s}|$.
The lack of majorization is consistent with the change of sign of $\Delta R_{\alpha}[\tilde{p}^{X,Z}]$ reported in Fig.~\ref{extvsint:fig} (solid line).

The other behaviors seen in Fig.~\ref{extvsint:fig} can be explained in the same way. For the products of marginals
 $\tilde{p}^{X}\tilde{p}^{Z}$  the four-dimensional ordered vectors for extreme and intermediate states are,
after Eqs. (\ref{ms}) and (\ref{mix}), , respectively:
$$\tilde{\lambda}^\prime = \tilde{\lambda} $$
and
$$\tilde{\mu}^\prime = \frac{1}{4} \left(  \left ( 1+\frac{|\bm{s} |}{2} \right )^2, 1- \frac{|\bm{s} |^2}{4}, 1- \frac{|\bm{s} |^2}{4}, \left  ( 1-
\frac{|\bm{s} |}{2} \right )^2 \right), $$
for balanced measurement.
Thus for all $|\bm{s} | \neq 0$ we have $\tilde{\mu}^\prime_1 > \tilde{\lambda}^\prime_1$ but  $\tilde{\mu}^\prime_1 + \tilde{\mu}^\prime_2
< \tilde{\lambda}^\prime_1 + \tilde{\lambda}^\prime_2 $, so that neither $\tilde{\lambda}^\prime \prec \tilde{\mu}^\prime$ nor $\tilde{\mu}^\prime \prec \tilde{\lambda}^\prime$. This correlates with the change of sign of the dashed line in Fig.~\ref{extvsint:fig}.

The same result is obtained for the comparison of the product of intrinsic statistics  $p^{X} p^{Z}$:
$$\lambda = \frac{1}{4} \left(1+ |\bm{s} |, 1+ |\bm{s} |,  1- |\bm{s} |, 1- |\bm{s} | \right) , $$
and
$$\mu = \frac{1}{4} \left( \left ( 1+\frac{|\bm{s} |}{\sqrt{2}} \right )^2, 1- \frac{|\bm{s} |^2}{2}, 1- \frac{|\bm{s} |^2}{2}, \left  ( 1-
\frac{|\bm{s} |}{\sqrt{2}} \right )^2 \right), $$
so that for all $|\bm{s} | \neq 0$ we get $\mu_1 > \lambda_1$ but $\mu_1 + \mu_2  < \lambda_1 + \lambda_2 $. This is consistent with the change of sign of the dotted line in in Fig.~\ref{extvsint:fig}.

Finally, in the three cases (joint, product of marginals and product of intrinsic statistics) the greatest component of the probability vector for the intermediate state is greater than the corresponding one for the extreme state. Consequently, for sufficiently large $\alpha$ the intermediate states provide the minimum, as seen in the three curves drawn in Fig.~\ref{extvsint:fig}. However a complete explanation of this figure cannot be provided by the lack of majorization relation.

\subsection{Comparison between joint and products of statistics}
\label{joint&product:sec}

It is well known that the R\'{e}nyi entropies do not fulfill in general the subadditivity property. Therefore for the same
state the entropy of the joint statistics $\tilde{p}^{X,Z}$ can be larger than the entropy of the product of its
marginals $\tilde{p}^{X} \tilde{p}^{Z}$. It can be easily checked that this is actually the case for intermediate
states.  This behavior implies that $\tilde{p}^{X,Z}$ and  $\tilde{p}^{X}\tilde{p}^{Z}$ can not be compared since
otherwise subadditivity will follow from Schur-concavity after the case $\alpha = 1$ where subadditivity holds.

Nevertheless, one may still wonder whether $\tilde{p}^{X,Z}$ and  $p^{X} p^{Z}$ are comparable or
not for intermediate states. We can easily show that there are cases for which they are comparable.

In this regard we can begin noting that $\tilde{p}^{X}  \prec p^{X}$ and  $\tilde{p}^{Z}  \prec p^{Z}$, since
these statistics are related through a doubly stochastic matrix
\begin{equation}
\pmatrix{\tilde{p}^X_{+} \cr \tilde{p}^X_{-} } =  \pmatrix{ \frac{1 + \eta}{2} & \frac{1- \eta}{2} \cr
\frac{1 - \eta}{2} & \frac{1 + \eta}{2} } \pmatrix{p^X_{+} \cr p^X_{-} }  ,
\end{equation}
and similarly replacing $X$ by $Z$, where $\eta = \cos \delta, \sin \delta$ for $X,Z$, respectively.  Notice that all
the matrix entries are positive and that each row and column sums to unity. 
However this does not imply any trivial relation between  $\tilde{p}^{X,Z}$ and  $p^{X} p^{Z}$. When comparing
the corresponding ordered distributions $\tilde{\mu}$ and $\mu$ for balanced joint measurements, 
we get $\tilde{\mu}_1 < \mu_1$ as well as  $\tilde{\mu}_1+ \tilde{\mu}_2 < \mu_1 + \mu_2$
for all $| \bm{s} | \neq 0$. However, we have $\tilde{\mu_1}+ \tilde{\mu}_2 + \tilde{\mu}_3 \leq \mu_1 + \mu_2
+  \mu_3$ for all $| \bm{s} | \leq 2 ( \sqrt{2} - 1) \approx 0.83$, while the opposite holds for
$| \bm{s} |$ above this value. This is to say, for intermediate states we get that the natural relation
$\tilde{p}^{X,Z} \prec p^{X} p^{Z}$ holds for mixed enough states with  $| \bm{s} | \leq 2 ( \sqrt{2} - 1)$,  while
otherwise the statistics are incomparable.

On the other hand, for extreme states we have always $\tilde{p}^{X,Z}  =  \tilde{p}^{X} \tilde{p}^{Z}
\prec p^{X} p^{Z}$.

\subsection{Majorization uncertainty relations}
\label{MajUrs:sec}

Majorization provides a rather neat form for uncertainty relations in terms of suitable constant vectors that majorize the statistics associated to the observables for every system state~\cite{HP,FGG}. In our case these are
\begin{equation}
\label{nur}
\tilde{p}^{X,Z} \prec \tilde{\omega}, \quad
\tilde{p}^X \tilde{p}^Z \prec \tilde{\omega}^\prime \quad \mbox{and} \quad
 p^X p^Z \prec \omega,
\end{equation}
where  $\tilde{\omega}$, $\tilde{\omega}^\prime$, and $\omega$ are constant vectors. By readily applying the procedure outlined in Ref.~\cite{FGG} we obtain 
\begin{eqnarray}
\label{mb}
 \tilde{\omega}& =& \frac{1}{4} \left(2, \sqrt{2} , 2 - \sqrt{2} , 0 \right) ,
\nonumber \\
 \tilde{\omega}^\prime &=& \frac{1}{16 \sqrt{2}} \left(9 \sqrt{2}, 8 - \sqrt{2}, 7 \sqrt{2} - 8, \sqrt{2} \right),
 \\
\omega &=& \frac{1}{8} \left(3+2\sqrt{2} , 5 - 2 \sqrt{2} , 0, 0\right) . \nonumber
\end{eqnarray}
Then corresponding uncertainty relations hold, for example for the joint distribution one has: \ $R_{\alpha} (\tilde{p}^{X,Z}) \geq R_{\alpha}(\tilde{\omega})$.

It is worth noting that there is a definite majorization relation between $\omega$ and the other two vectors, that is
\begin{equation}
\tilde{\omega} \prec \omega \quad \mbox{and} \quad \tilde{\omega}^\prime \prec \omega.
\end{equation}
These two relations are quite natural and express that the uncertainty lower bound is larger for the statistics derived from simultaneous joint measurement, either $\tilde{p}^{X,Z}$ or $\tilde{p}^X \tilde{p}^Z$, than for the exact intrinsic statistics. This is the majorization relation counterpart of the
well-known result that the variance-based lower bound for operational position--momentum uncertainty is at least four times the intrinsic one~\cite{AG}.

However, there is no majorization relation between $\tilde{\omega}$ and $\tilde{\omega}^\prime$ since while $\tilde{\omega}_1^\prime > \tilde{\omega}_1$, we have  $\tilde{\omega}_1^\prime + \tilde{\omega}_2^\prime + \tilde{\omega}_3^\prime < \tilde{\omega}_1 + \tilde{\omega}_2 + \tilde{\omega}_3$.

Finally we show that there are no system states leading to statistics equating the distributions~(\ref{mb}).
To this end we use Eqs.~(\ref{mis}), (\ref{js}), and~(\ref{ms}) to determine the values of $s_x$ and $s_z$ that would lead to $\tilde{p}^{X,Z}$, $\tilde{p}^X \tilde{p}^Z$
and $p^X p^Z$, equating $\tilde{\omega}$, $\tilde{\omega}^\prime$ and $\omega$, respectively. Without loss of generality we consider $s_x$ and $s_z$ to be positive. For the joint statistics, the null component in $\tilde{\omega}$ implies that $s_x = s_z = \frac{1}{\sqrt{2}}$. Thus according to~(\ref{js}) the other values for $\tilde{p}^{X,Z}$ should be $\frac 12$, $\frac 14$ and $\frac 14$, which are not equal to the corresponding values in $\tilde{\omega}$. For the product of marginals $\tilde{p}^X \tilde{p}^Z$, the sum of the greatest and lowest components of $\tilde{\omega}^\prime$ imply that $s_x = s_z = \frac{1}{\sqrt{2}}$. Thus after~(\ref{ms}) the other two components for $\tilde{p}^X \tilde{p}^Z$ should be both $\frac{3}{16}$, which are not equal to the corresponding values in $\tilde{\omega}^\prime$. For the intrinsic statistics, we have that the two zeros of $\omega$ imply that either $s_x=0$ or $s_z=0$. In any case~(\ref{mis}) would then imply that the other components of $p^X p^Z$ should be both $\frac 12$, which is different from the corresponding values in $\omega$.

\section{Duality relation}
\label{duality:sec}

Following the approach in Ref.~\cite{BPHP} we may compare these entropic results
with some other assessments of joint uncertainty or complementarity. Among them,
one of the most studied is the duality relation between path knowledge and visibility
of interference in a Mach--Zehnder interferometric setting~\cite{duality1,duality2}. This fits with
our approach by regarding $|\pm \rangle$ as representing the internal paths of a
Mach--Zehnder interferometer, while $| a_\pm \rangle$ represent the states of the
apparatus monitoring the path followed by the interfering particle.

One of most used duality expression is~\cite{duality1}
\begin{equation}
D^2 + V^2 \leq 1 ,
\end{equation}
where $D = \mathrm{Tr}_A \left ( \left | w_+ \rho_A^{(+)} - w_- \rho_A^{(-)}
\right | \right )$ is the so-called distinguishability. Regarding  the particular
case where the system and apparatus are in pure states, we have
$\rho_A^{(\pm)} = | a_\pm \rangle \langle a_\pm |$ and $w_+ = 1 -w_- = \cos^2
\frac{\theta}{2}$ , so that
\begin{equation}
\label{D}
D = \sqrt{1 - 4 w_+ w_- \left | \langle a_+ | a_- \rangle \right |^2 } .
\end{equation}
This represents the knowledge available about the path followed by the particle, which
 is {\it grosso modo} inversely proportional to path uncertainty. On the other hand,
the interference is assessed by the standard fringe visibility $V$ obtained
when the relative phase $\varphi$ is varied in Eq.~(\ref{pss}),
\begin{equation}
\label{V}
V = 2 \sqrt{w_+ w_-} \, \left | \langle a_+ | a_- \rangle \right | .
\end{equation}
This roughly speaking represents the phase uncertainty, the counterpart of the
uncertainty of $\sigma_x$ in our approach. Note that in these duality relations
path and interference are not treated symmetrically, contrary to the approach developed
here in terms of entropic measures.

After Eqs.~(\ref{D}) and~(\ref{V}) we can appreciate that $D^2 + V^2
= 1$ whenever the system and apparatus are in pure states. This is to say that this duality
relation is blind to the differences between extreme and intermediate states, in sharp contrast to
the more complete picture provided by the entropic measures with equal entropic indices.
This was already shown in Ref.~\cite{BPHP} regarding its intrinsic counterpart $P^2 + V^2
\leq 1$, where $P = | w_+ - w_- |$ is the predictability. Nevertheless, an equivalence with the
duality relation is obtained, using conjugated entropic indices that lead to the so-called
min--max entropies, as was recently shown in Ref.~\cite{Coles2014}.

Since the duality relation does not discriminate between pure states it may be interesting
to complete the duality analysis by examining the states of maximum $D$ or
$V$, as well as those states with $D = V$.

{}From Eq.~(\ref{D}) the maximum distinguishability, $D = 1$, holds either when $w_+ =0$, $w_- =0$,
or $\langle a_+ | a_- \rangle=0$. These are all the cases where the particle actually
follows just a single path, or when the apparatus can provide full information about the
path followed. On the other hand, after Eq.~(\ref{V}), the
maximum visibility, $V =| \langle a_+ | a_- \rangle |$, holds when
both paths are equally probable $w_+ = w_- = \frac 12$. Furthermore the maximum visibility
reaches unity, $V =1$ when $| a_+ \rangle$ is proportional to $| a_- \rangle$. This is when
both paths are equally probable and the apparatus provides no information about the path.
Within the set $S$, the extreme states $s_z=\pm 1$ satisfy the requirements for extreme
distinguishability, while those with $s_x = \pm 1$ reach maximum visibility. This agrees
with the case of unobserved duality~\cite{BPHP}.

On the other hand, $D = V$ holds provided that $w_+ w_- \, | \langle a_+
| a_- \rangle |^2 = \frac 18$. For balanced detection, $| \langle a_+ | a_- \rangle | = \frac{1}{ \sqrt{2}}$
so that $w_+ w_- =\frac 14$ and then $w_+ = w_- = \frac 12$. Within the set $S$ this is satisfied
by the extreme states being eigenstates of $\sigma_x$. Contrary to what happens for the
unobserved duality relation, the intermediate states do not satisfy $D
= V$.
The extreme $s_x = \pm 1$ can reach both maximum visibility and $D = V$ since for
balanced joint detection we get $D \geq \frac{1}{\sqrt{2}} \geq V$ for all states.

\section{Concluding remarks}
\label{conclu:sec}

We have presented several examples of application of R\'enyi entropies as measures of quantum uncertainty in the case of simultaneous measurements. We have explored those situations leading to unexpected or contradicting predictions for different entropies
and states as reported in Sec.~\ref{intvsextr:sec}. We have shown that the interplay  between extreme and intermediate states as those of minimal uncertainty, shown in Figs.~\ref{Rentvstheta:fig} and~\ref{extvsint:fig}, is consistent with lack of majorization relation between the corresponding statistics, as we have discussed in Sec.~\ref{ext&int maj:sec}.

Moreover, we have compared the joint and products of statistics in connection with majorization in Sec.~\ref{joint&product:sec}. We have obtained that for the intermediate states, the joint distribution is majorized by the product of intrinsic statistics up to certain degree of purity.
On the other hand, for extreme states this situation holds for any degree of purity as naturally one could have expected.

In Sec.~\ref{MajUrs:sec} we have obtained the corresponding majorization uncertainty relations for the joint, product of marginal and product of  intrinsic statistics, obtaining the corresponding majorizing constant distributions $\tilde{\omega}$, $\tilde{\omega}^\prime$ and $\omega$. We have seen that there exist majorization relations between $\omega$ and $\tilde{\omega}$, and between $\omega$ and $\tilde{\omega}^\prime$. This means that the uncertainty lower bound is larger for the statistics derived from simultaneous joint measurement, either $\tilde{p}^{X,Z}$ or $\tilde{p}^X \tilde{p}^Z$, than for the exact intrinsic statistics. This can be interpreted as the majorization relation counterpart of the well-known result that the variance-based lower bound for operational position--momentum uncertainty is at least four times the intrinsic one. In addition, we have proved that these majorization uncertainty relations are not tight, in the sense that there is no state
reaching the bounds.

In Sec.\ref{duality:sec}, we have shown that the new uncertainty relations are much more comprehensive than the traditional assessment
of complementarity in terms of distinguishability $D$ and visibility $V$. For a more fruitful comparison we have developed
the traditional approach inquiring about the states with extreme $D$ or $V$ , as well as about the intermediate states $D=V$.

The results presented in this work intend to provide a better understanding of uncertainty relations. In recent times there has been a growing interest in applying advanced
statistical tools to quantum problems, going beyond the simple use of variances or entropies.
Thus, majorization emerges as a powerful tool to understand fundamental aspects of quantum uncertainty and complementarity in the most complete and simple form.

\section*{Acknowledgments}

A.L.\ acknowledges support from Facultad de Ciencias Exactas de la Universidad Nacional de
La Plata (Argentina), as well as Programa de Intercambio Profesores de la Universidad Complutense and projects
FIS2012-35583 of spanish Mi\-nis\-te\-rio de Econom\'{\i}a y Competitividad and Comunidad Aut\'onoma de Madrid
research consortium QUITE\-MAD+ S2013/ICE-2801.
G.M.B. and M.P. acknowledge support from CONICET (Argentina). 
M.P.\ is also grateful to R\'egion Rh\^{o}ne--Alpes (France) and Universidad Complutense de Madrid (Spain).
A.L.\ acknowledges Th.\ Seligman for valuable suggestions. We thank J.\ Kaniewski for valuable comments.

\end{document}